\documentclass[letterpaper,10pt]{article} 
\usepackage{opticameet3}

\usepackage{amsmath,amssymb}
\usepackage{upgreek}
\usepackage{multirow}
\usepackage{pgfplots}
\pgfplotsset{compat=1.3} 
\usepackage{graphicx}
\usepackage{subcaption}
\usepackage{wrapfig, blindtext}
\usepackage{mwe}
\usepackage{adjustbox}
\usepackage{xcolor}
\usepackage{collcell}
\usepackage{multirow}
\usepackage{mathtools}
\usepackage[colorlinks=true,bookmarks=false,citecolor=blue,urlcolor=blue]{hyperref} 
\usetikzlibrary{arrows, arrows.meta}

\begin{document}

\title{ Knowledge Distillation Applied to Optical Channel Equalization: Solving the Parallelization Problem of Recurrent Connection}
\author{Sasipim Srivallapanondh\textsuperscript{(1)}, Pedro J. Freire\textsuperscript{(1)}, Bernhard Spinnler\textsuperscript{(2)}, Nelson Costa\textsuperscript{(3)}, Antonio Napoli\textsuperscript{(2)}, Sergei K. Turitsyn\textsuperscript{(1)}, Jaroslaw E. Prilepsky \textsuperscript{(1)}}
\address{\textsuperscript{(1)}Aston University, Birmingham, UK; \hspace{0.5mm}
   \textsuperscript{(2)}Infinera, Munich, Germany; \hspace{0.5mm} \textsuperscript{(3)}Infinera, Carnaxide, Portugal

   }
  \vspace{-0.5mm}
\email{s.srivallapanondh@aston.ac.uk}

\copyrightyear{2022}

\vspace{-1mm}
\begin{abstract}
To circumvent the non-parallelizability of recurrent neural network-based equalizers, we propose \textit{knowledge distillation} to recast the RNN into a parallelizable feedforward structure. The latter shows 38\% latency decrease, while impacting the Q-factor by only 0.5dB.
\end{abstract}

\vspace{-0.4mm}
\section{Introduction}
\vspace{-1.5mm}
Optical fiber nonlinearity significantly limits the information rate in current coherent transmission systems. Moreover, with the ever-increasing transmission bandwidth, nonlinearity becomes even more important \cite{xu2021information}. Various digital signal processing (DSP) techniques have been proposed to minimize nonlinear effects \cite{cartledge2017digital}. Due to the universal approximation capability of neural networks (NNs), the NNs have recently been intensively studied for the optical channel post-equalization, because they can approximate the inverse optical channel transfer function with good accuracy and revert the nonlinear distortions. 
In particular, recurrent NNs (RNN) based equalizers have shown the best capability in equalizing nonlinear impairments as compared to the feed-forward NN types \cite{Pedro2021,deligiannidis2020compensation, deligiannidis2021performance}. However, since the RNN structure has a feedback loop, it is not easily parallelizable. This characteristic makes it challenging to implement the RNNs in low-complexity hardware for high-speed processing \cite{chang2017hardware}. Indeed, parallelization is the key to achieving low latency and, simultaneously, high throughput required by high-speed optical networks, as parallel computing increases the throughput and reduces the computational time \cite{robey2021parallel}.
To enable parallelization, in this work, we adopt the concept of knowledge distillation (KD) to transform the RNN-based equalizer into a feedforward structure. Generally, KD is a method to transfer the knowledge from a larger model, known as a \emph{teacher} model, to a more compact one (a \emph{student}), which requires fewer computations \cite{hinton2015distilling}. Most of the prior works using KD have focused on classification tasks where the teacher and student have a similar topology. However, KD for a regression task and the cross-architecture KD, as we propose in our work, have only been studied very recently \cite{xu2022contrastive}. In this work, we also reduce the computational complexity (CC) per recovered symbol by recovering a multi-symbol output \cite{sang2022low}.

In this paper, for the first time, we present a workaround to solve the RNN parallelization issue, by using KD. The KD, in our case, is used to transfer the knowledge from a recurrent teacher model which is a bidirectional long-short term memory (biLSTM) coupled with 1D-convolutional NN (1D-CNN), the efficient equalizer model proposed in \cite{freire2022reducing}, to a feedforward student model: a 1D-CNN, shown in Fig.~\ref{fig:parallelization}(a). Furthermore, we point out that together with the realization of a parallelizable structure, we reduce the inference latency of the model obtained via KD by 38\%. Finally, we also observed a slight 0.5~dB reduction in Q-factor at optimum power.

\vspace{-1.5mm}
\section{Parallelization of Recurrent Equalizers via Knowledge Distillation}
\begin{figure}[htbp]
\centering
  \includegraphics[scale=0.30]{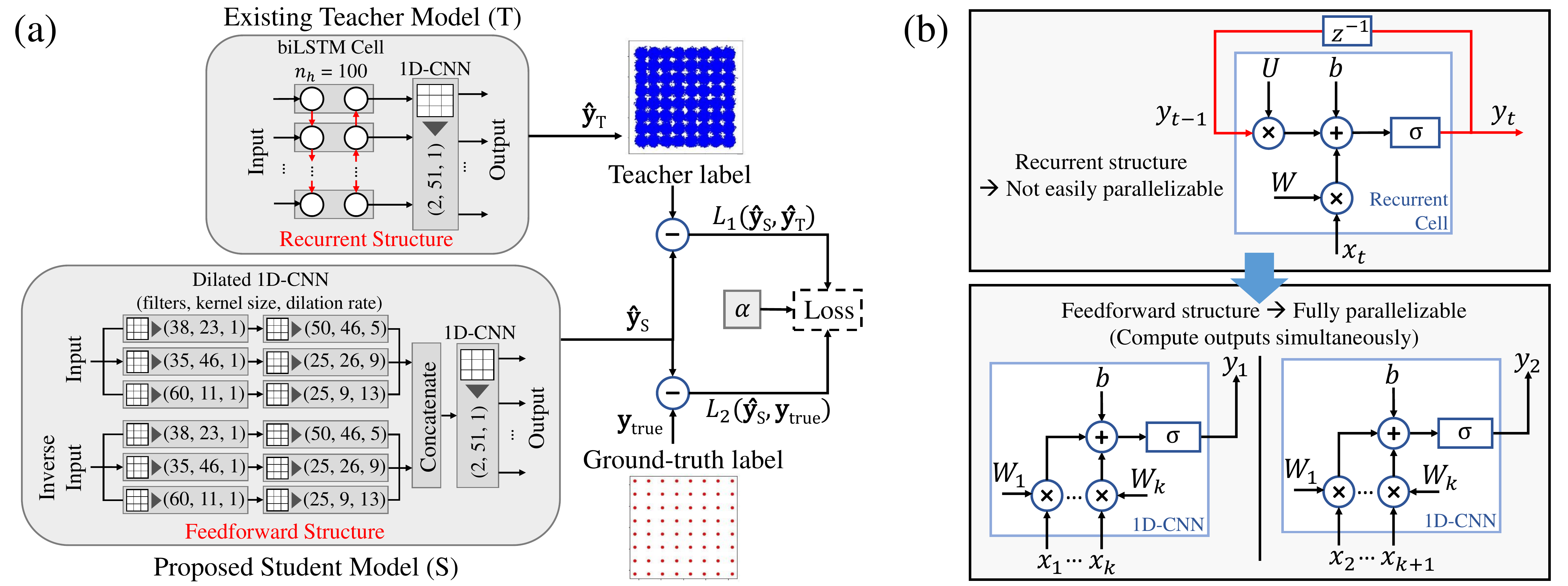}
  \captionof{figure}{(a) KD framework with biLSTM+1D-CNN as a teacher model and dilated 1D-CNN as a student model; (b) Illustration of the parallelizability comparison between a recurrent cell (top) 
  and a 1D-CNN (bottom).} 
  \label{fig:parallelization}
  \vspace{-6mm}
\end{figure}

\vspace{-1.5mm}
The recurrent nature of biLSTM, which takes into account the input of the current stage and the output of the previous stage, makes it very useful for learning sequential data. Furthermore, the input flowing in both directions further enhances the biLSTM learning. Due to the sequential processing of the double-recurrent setting and, consequently, limited parallelizability, the RNN-based model is not easily implementable for high-speed transmission. On the other hand, 1D-CNNs are feedforward-based NNs where the input temporal sequential batches are processed independently, hence, parallel operations are possible. Fig.~\ref{fig:parallelization}(b) illustrates the recurrent structure at the top with a feedback loop, preventing parallelization, while the feedforward structure at the bottom can process multiple sets of inputs and provide multiple outputs simultaneously. Transforming the model architecture from the biLSTM to 1D-convolutional layers enables parallel computation for the previously proposed recurrent structure-based equalizer (NNE) or the teacher model.

Fig.~\ref{fig:parallelization}(a) shows the teacher and student model structures and the KD process. The teacher NN contains a biLSTM layer with 100 hidden units and a linear 1D-CNN layer with two filters and a kernel size of 51. The student model has a dilated convolutional structure, which allows the NN to deal with long-term temporal dependencies and have large receptive fields within only a few layers \cite{oord2016wavenet,xu2022contrastive}. The last 1D-CNN layer of both the teacher and the student has the same parameters. To mimic biLSTM, which learns the input data in forward and backward directions, the student model learns both directions of the training data. The backward direction input means the input sequence (forward direction) in reverse time order. The Bayesian Optimizer (BO) \cite{Pedro2021} is used to optimize the hyperparameter values of the student model. The estimated optimal values are depicted in Fig.~\ref{fig:parallelization}(a). Note that (38, 23, 1) means that the 1D-CNN layer operates with 38 filters, a kernel size of 23, and a dilation rate of 1. The activation function of the dilated 1D-CNN part is LeakyRelu. The KD loss approach (Eq.~(\ref{eq.kdloss})) for the student model training implies a joint loss function that takes into account both the loss between the student's and teacher's predictions and the loss between the student's predictions and ``ground truth'':
\vspace{-1mm}
\begin{equation}\label{eq.kdloss}
\begin{gathered}
L_{\text{KD}}= \alpha L_1(\boldsymbol{\hat{y}}_S,\boldsymbol{\hat{y}}_\text{T}) + (1-\alpha) L_2(\boldsymbol{\hat{y}}_S,\boldsymbol{y}_\text{true}),
    \end{gathered}
\vspace{-1mm}
\end{equation}
where $\boldsymbol{y}_\text{true}$ is the ground-truth labels, $\boldsymbol{\hat{y}}_S$ and $\boldsymbol{\hat{y}}_\text{T}$ represent the student's and the teacher's predictions, respectively, and $\alpha$ is the hyperparameter balancing the contribution of each term to the final loss. The teacher model is pre-trained and used only to create teacher labels. The student model's training with KD is carried out for 1000 epochs, mini-batch size of $2^{10}$; the learning rate and $\alpha$ found by BO are equal to 0.00026 and 0.903, respectively. The dataset is obtained by numerical simulation, assuming the transmission of a single 30~GBd, 64-QAM  dual-polarization channel along $20\times50$~km standard single mode fiber (SSMF) spans. At the end of each fiber span, optical fiber losses are compensated for by an Erbium-Doped Fiber Amplifier (EDFA) with a 4.5~dB noise figure.

Typically, the NNEs proposed in previous works are designed to recover one symbol at a time \cite{Pedro2021}. However, the single-symbol output NNEs can be computationally inefficient, as the weights and biases trained to recover one symbol may still be useful for recovering multiple symbols \cite{sang2022low}. 
Therefore, we focus on multi-symbol output equalizers to reduce the CC per recovered symbol. The last layer of our model adopts the 1D-CNN layer containing two filters with a linear activation function to recover both real and imaginary parts of the signal. A set of 221 input symbols is fed to the NN simultaneously, to recover 171 symbols at the output at each inference step.

\vspace{-1.5mm}
\section{Results and Discussion}
\begin{figure*}[ht!] 
  \vspace{-4mm}
\begin{subfigure}{.32\textwidth}
    \centering
 \begin{tikzpicture}[scale=0.58]
    \begin{axis} [ 
        xlabel={Launch power [dBm]},
        ylabel={Q-Factor [dB]},
        grid=both,  
        xmin=-3, xmax=5,
    	xtick={-3, ..., 5},
    	ymin=1.7, ymax=11,
        legend style={legend pos=south west, legend cell align=left,fill=white, fill opacity=0.6, draw opacity=1,text opacity=1},
    	grid style={dashed}]
        ]
    \addplot[color=blue, mark=square, very thick]   
    coordinates {
    (-3, 7.88)(-2, 8.69)(-1, 9.49)(0, 10.1)(1,10.5)(2,10.66)(3,10.37)(4,9.7)(5,8.6)
    };
    \addlegendentry{Teacher model (biLSTM+CNN)};
    
    \addplot[color=red, mark=*, very thick]   
    coordinates {
    (-3, 7.856)(-2, 8.66)(-1,9.35)(0, 9.88)(1, 10.196)(2, 10.19)(3, 9.6)(4, 8.6)(5, 7.5)
    };
    \addlegendentry{Student model with KD (1D CNN)};
        \addplot[color=green, mark=o, very thick, dash pattern={on 7pt off 2pt}]   
    coordinates {
    (-3,7.01)(-2,7.30)(-1, 7.52)(0, 7.87)(1,8.12)(2,8.3)(3, 8.23)(4, 7.61)(5, 6.52)
    };
    \addlegendentry{Student model trained from scrath};
    
    \addplot[color=cyan, mark=x, very thick, dash pattern={on 7pt off 2pt}]   
    coordinates {
    (-3,7.44)(-2, 8.00)(-1, 8.57)(0, 8.95)(1, 9.19)(2,9.04)(3,8.80)(4, 8.01)(5, 6.83)
    };
    \addlegendentry{Student model with L2 reg.};
    
    \addplot[color=orange, mark=+,very thick, dotted]    coordinates {
    (-3,  7.79)(-2, 8.46)(-1, 8.96)(0, 9.2)(1,9.1)(2,8.6)(3,7.7)(4,6.5)(5,5.1)
    };
    \addlegendentry{DBP 1 STpS};
    
    \addplot[color=black,mark=triangle, very thick, dotted]    coordinates {
    (-3, 7.4)(-2, 7.7)(-1, 7.8)(0, 7.3)(1,6.6)(2,5.6)(3,4.4)(4,3)(5,1.7)
    };
    \addlegendentry{CDC (Regular DSP)};

    \end{axis}
    \draw[ <->, red] (2, 5.18) -- (2,5.51);
    \draw [dashed, red] (1.7,5.18) -- (5.2,5.18);
    \draw [dashed, red] (1.7,5.51) -- (5.2,5.51);
    \node[text width=1cm, red] at (0.95,5.35) 
    {0.5dB};
    \end{tikzpicture}
\end{subfigure}  
\hfill
\begin{subfigure}{.31\textwidth}
    \centering
 \begin{tikzpicture}[scale=0.572]
     \begin{axis}[
    ybar=5,
    symbolic x coords={CPU, GPU},
    bar width=0.4cm,
    enlarge x limits=0.4,
        ymin=0.0002,
    xtick=data,
    ymajorgrids = true,
    ylabel style={align=center},
    ylabel = {Inference time per recovered symbol (sec)\\Result obtained on Google Colab} ,
    xlabel = {Inference engine},
    legend style={nodes={scale=1, transform shape},legend pos=north east, legend cell align=left,fill=white, fill opacity=0.6, draw opacity=1,text opacity=1},
    ]
      \addplot[style={blue,fill=blue,mark=none}]
            coordinates {(CPU, 0.000406) (GPU, 0.0003561)};
    \addplot[style={red,fill=red,mark=none}]
          coordinates {(CPU, 0.0003356) (GPU, 0.000221)};
    \draw[ <->, black] (7.5, 135) -- (7.5,206);
    \draw [dashed, black] (-20,135) -- (20,135);
    \draw [dashed, black] (-20,206) -- (20,206);
    \node[text width=1cm, black, thick] at (27,170) 
    {17\%};
              
    \draw[ <->, black] (107.5, 21) -- (107.5,156);
    \draw [dashed, black] (80,21) -- (120,21);
    \draw [dashed, black] (80,156) -- (120,156);
    \node[text width=1cm, black, thick] at (125,83) 
    {38\%};

    \draw[>=latex, thick, <->, red] (32, 165) -- (82,83);
        \node[scale=1.5, red, thick] at (47,110) 
    {2.3X};
    \legend {Teacher model, Student model with KD};
  \end{axis}
    \end{tikzpicture}
\end{subfigure}\hfill
\includegraphics[scale=0.28]{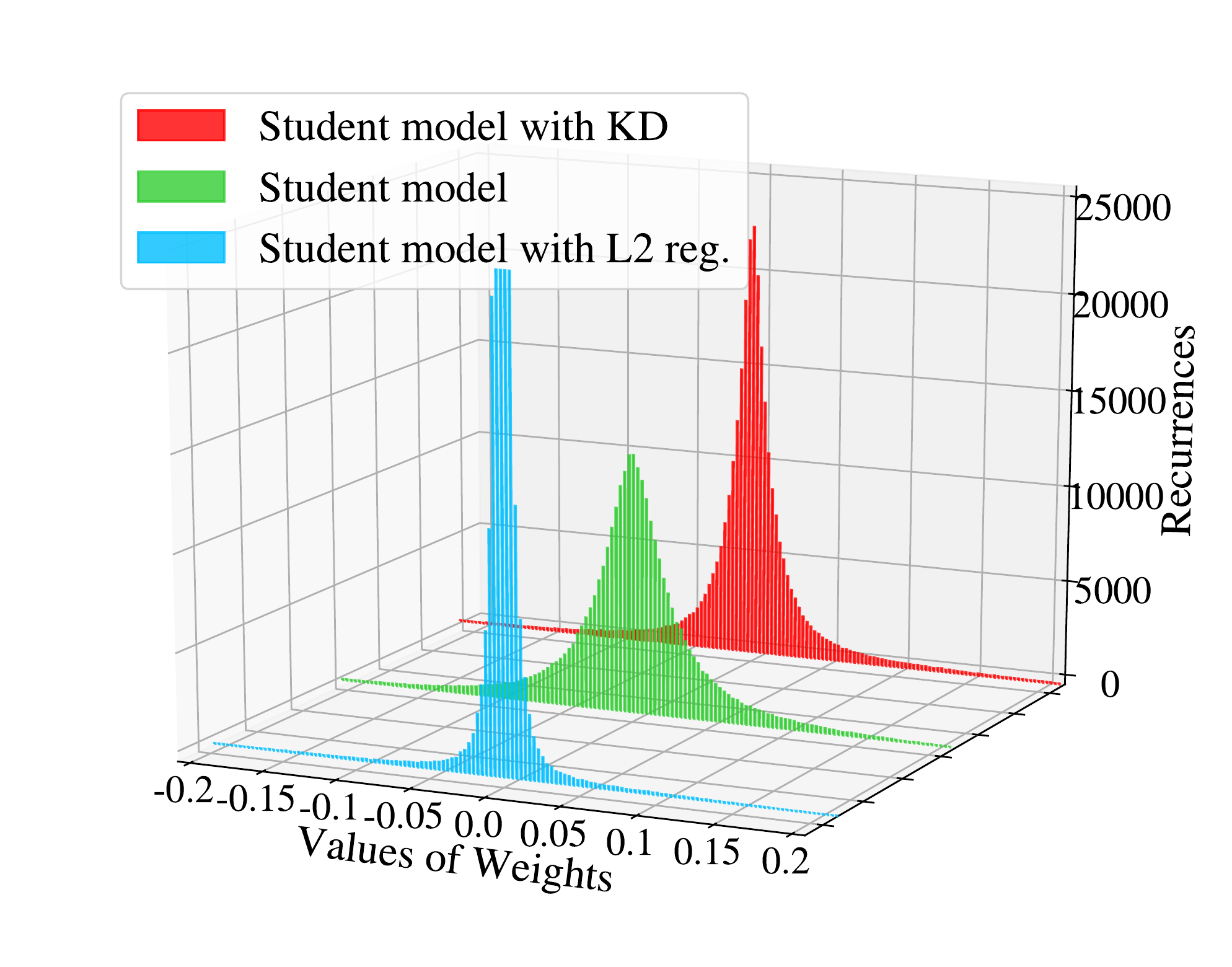}
  \put(-445, 112){\textcolor{black}{(a)}}
  \put(-290, 112){\textcolor{black}{(b)}}
  \put(-158, 112){\textcolor{black}{(c)}}
\vspace{-2mm}
  \caption{(a) Q-factor as a function of the launch power for the NNE obtained via KD, compared to the original (teacher) model, CDC, and 1~STpS DBP; (b) Inference time of teacher/student models; (c) Weight distribution of student model trained with different approaches.}
  \label{fig:results2} 
  \vspace{-6mm}
\end{figure*}
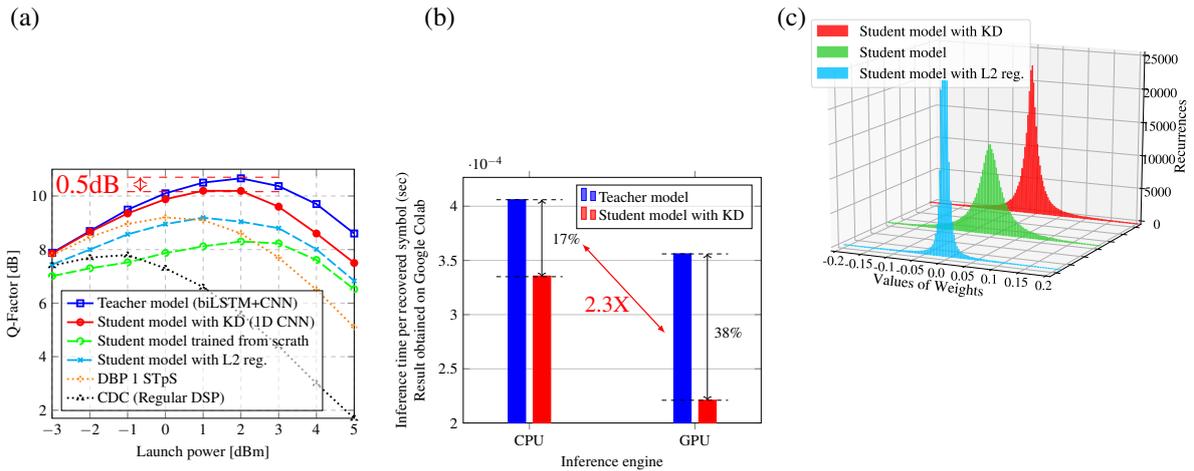

\vspace{-1.5mm} 
We compare our proposed student model (1D-CNN) using the KD framework with the teacher model (biLSTM +CNN) pre-trained as in \cite{freire2022reducing}, the student model trained from scratch (without KD) with exactly the same settings, and the student model trained from scratch with L2 regularizer \cite{ng2004feature}. The optimum L2 coefficient depends on the launch power. At 2~dBm launch power, the optimum L2 coefficient found by grid search is $10^{-4}$. Fig.~\ref{fig:results2}(a) depicts Q-factor vs. launch power for different types of NNEs. We can observe that the Q-factor performance of the feedforward student model with KD drops by 0.5~dB compared to the recurrent teacher model at its optimal launch power (2~dBm). With KD, the performance of the student model is comparable to that of the teacher model in the linear transmission regime, but the student's performance degrades slightly as the launch power increases. However, when training the student model from scratch without KD, the model suffers from overfitting, resulting in a noticeable degradation of the peak performance by 2.4~dB at its optimal power. Training the student model with the L2 regularizer, which helps enhance the generalization capability, improves the performance of the NN compared to training it from scratch only, but still does not reach a similar performance level as the one achieved when the student model is trained with KD. The performance achieved using digital back propagation (DBP) 1 step/span (STpS) and chromatic dispersion compensation (CDC) are also shown for reference. 

Although the teacher model leads to slightly better optical performance than the student model with KD, Fig.~\ref{fig:results2}(b) shows a reduction in the inference time per recovered symbol of the student model with KD of 17\% and 38\% when using the CPU and GPU, respectively. Note that both the teacher and the student models recover 171 symbols per inference step. Using a GPU leads to a more significant reduction in latency because of its better parallel-computing ability, which enables it to better exploit the parallelizability of the student feedforward structure.
The parallelization of the proposed feedforward equalizer and its savings in latency are key to the real-time hardware implementation of NNEs. 

Now, we study the features associated with the KD-trained model. For this purpose, we also report the weight distribution of the student model trained with different approaches in  Fig.~\ref{fig:results2}(c). Compared to the student model trained from scratch, the student model with KD has a more regularized weight distribution: the weights are more concentrated around zero. This characteristic helps reduce the model's variance and overfitting. The optimal value of $\alpha$ in the KD loss function is 0.903, which means that the student model learns 90.3\% from the teacher labels and the rest comes from the ground-truth labels. 
This fact demonstrates the effectiveness of the teacher labels in the student's learning. The teacher constellation/labels depicted in Fig.~\ref{fig:parallelization}(a) show that the teacher also provides helpful information on the noise, whereas this information cannot be encoded in the ground-truth labels (which contain only real values). The weight distribution of the student model with KD and the improvement in Q-factor, compared to the training of the 1D-CNN without KD, both support the concept of using teacher labels as efficient regularizers \cite{hinton2015distilling}.

\vspace{-1.5mm}
\section{Conclusions}
\vspace{-1mm}
In this work, for the first time, the knowledge distillation technique has been proposed as an efficient tool to achieve the parallelizability of the recurrent equalizers: in our study, KD transfers the knowledge from a recurrent-connection-based biLSTM equalizer to a parallelizable feed-forward 1D-CNN. This approach enables the parallelization of signal processing, allowing us to essentially simplify the hardware implementation of NN models. We also show that the proposed feedforward equalizer obtained with KD, results in a reduced signal processing latency by 38\% compared to the original biLSTM model, at the expense of slightly reducing the maximum Q-factor by 0.5~dB.\\

\vspace{-1.5mm}
\footnotesize
\linespread{0.0}
\textbf{Acknowledgements}: This work has received funding from the EU Horizon 2020 program under the Marie Sk\l{}odowska-Curie grant agreement No.~956713 (MENTOR) and 813144 (REAL-NET). SKT acknowledges the support of the EPSRC project TRANSNET (EP/R035342/1).
\normalsize
\linespread{1.0}

\end{document}